\documentstyle[preprint,aps,epsfig]{revtex}
\begin{document}
\draft
\preprint{\vbox{\hbox{IFT--P.065/97}}}
%
\title{Anomalous Higgs Couplings in Triple Gauge Boson Production at the NLC
{\footnote{To appear in the  Proceedings of the {\it ECFA/DESY Study on Physics and 
Detectors for the Linear Collider}, report DESY 97-123E.}}}
\author{F.\ de Campos, S.\ M.\ Lietti, S.\ F.\ Novaes and R.\ Rosenfeld}
\address{Instituto de F\'{\i}sica Te\'orica, 
Universidade  Estadual Paulista, \\  
Rua Pamplona 145, CEP 01405--900 S\~ao Paulo, Brazil.}
\date{\today}
\maketitle
\widetext
\begin{abstract}
We present the sensitivity limits on the coefficients of a
dimension--6 effective operators that parametrizes the possible
effects of new physics beyond the Standard Model. Our results are
based on the study of the processes $e^+ e^- \to W^+ W^- \gamma$,
$Z Z\gamma$, and $Z\gamma \gamma$ at NLC energies. In our
calculations, we include {\it all} the anomalous interactions 
involving vector and Higgs bosons, and take into account the
Standard Model irreducible background. We analyse the impact of
these new interactions on the total cross section and on some
kinematical distributions of the  final state particles.
\end{abstract}


\section{Introduction} 
\label{int}

One of the main physics goals of LEP2 and future $e^+ e^-$
colliders is to directly test the gauge nature of couplings among
the electroweak gauge bosons. Deviations from the Standard Model
(SM) predictions for these couplings would indicate the existence
of new physics effects. The most general phenomenological
parametrization for these  deviations can be achieved by means of
an effective Lagrangian \cite{effective} that involves, apart
from the SM Lagrangian, high--dimension operators that describe
the new phenomena, containing the relevant fields at low energies
and respecting the symmetries of the Standard Model.

The effective Lagrangian approach is a model--independent way to
describe  new physics that can occur at an energy scale $\Lambda$
much larger than the scale where the experiments are performed
\cite{param}.  Here we focus on a linearly realized $SU_L(2)
\times U_Y(1)$ invariant  effective Lagrangian to describe the
bosonic sector of the Standard Model,  keeping the fermionic
couplings unchanged. In order to write down the  most general
dimension--6 effective Lagrangian containing all SM  bosonic
fields, {\it i.e.\/} $\gamma$, $W^{\pm}$, $Z^0$, and $H$,  we
adopt the notation of Hagiwara {\it et al.} \cite{hisz}. This
Lagrangian has eleven independent operators in the linear
representation that are locally $SU_L(2) \times U_Y(1)$
invariant,  $C$ and $P$ even. We discard the four operators which
affect the  gauge boson two--point functions at tree--level and
therefore are  strongly constrained by LEP1 measurements. We also
do not consider  the two operators that modify only the Higgs
boson self--interactions, since they  are not relevant for our
calculations. We are then left with five independent operators,
and the Lagrangian is written as,
\begin{equation}
{\cal L}_{\text{eff}} = {\cal L}_{\text{SM}} + 
\frac{1}{\Lambda^2} \left(  
f_{WWW} {\cal O}_{WWW} + f_{WW} {\cal O}_{WW} + f_{BB} {\cal O}_{BB} +
f_W {\cal O}_{W} +f_B {\cal O}_{B} \right)
\; , 
\label{lagrangian}
\end{equation}
with each operator ${\cal O}_i$ defined as in ref.~\cite{hisz}.
It is important to notice that this effective Lagrangian, often
used to describe anomalous trilinear gauge couplings, can  lead
to anomalous quartic interaction among gauge bosons and also to
anomalous couplings of these particles with the Higgs field. The
operator ${\cal O}_{WWW}$ contributes only to anomalous gauge
boson self couplings ($VVV$, $VVVV$), ${\cal O}_{WW}$ and ${\cal
O}_{BB}$  contribute only to anomalous Higgs couplings ($HVV$),
whereas ${\cal O}_{W}$ and ${\cal O}_{B}$ give rise to both kinds
of new couplings.  All these interactions should be investigated
at the NLC  in order to  search for hints about the nature of the
new physics described  by these higher dimensional operators.

In this contribution we present the results of our analyses
constraining the coefficients of the dimension--6 effective
Lagrangian (\ref{lagrangian}) via the processes  $e^+ e^- \to W^+
W^- \gamma$ \cite{us}, $Z Z \gamma$,  and $Z \gamma \gamma$
\cite{us2} at the  Next Linear Collider (NLC). Some related works
include the references \cite{quartic,genuine,hagiwara2,ee,our,gamma}. 

We chose to study the reaction $e^+ e^- \to W^+ W^- \gamma$ since
it is the process with the largest cross section involving
triple, quartic gauge boson couplings and also anomalous
Higgs--gauge boson couplings. Therefore, it is also sensitive to
$f_{WW}$ and $f_{BB}$, offering an excellent possibility for a
detailed study of these couplings. It is important to notice that
studies of anomalous trilinear gauge boson couplings from
$W$--pair production will significantly constrain combinations of
the parameters $f_{WWW}$, $f_W$ and $f_B$. However, they are
``blind" with respect to $f_{WW}$ and $f_{BB}$.

Another way to test the anomalous Higgs couplings generated by
the operators ${\cal O}_{WW}$ and ${\cal O}_{BB}$ is through the
analysis of the processes $e^+ e^- \to Z Z \gamma$ and $Z \gamma
\gamma$, that are only sensitive to $H Z \gamma$ and $H \gamma
\gamma$ anomalous Higgs interactions which, in the SM,  appear
only at one--loop level \cite{hgg,hgz}. 
 
\section{Anomalous $WW\gamma$ Production at NLC}

The Standard Model cross section for the process $e^+ e^- \to W^+
W^- \gamma$ was evaluated in Ref.\ \cite{SM}. Including the new
contributions from (\ref{lagrangian}),  there are 16 standard and
26 anomalous Feynman diagrams involved in the reaction $e^+ e^-
\to W^+ W^- \gamma$ \cite{foot}. In order to compute these
contributions, we have incorporated all new  couplings in a {\tt
Helas}--type \cite{helas} Fortran subroutines. These new
subroutines were used to extend a {\tt Madgraph} \cite{madgraph}
generated  code to include all the anomalous contributions and to
numerically evaluate  the helicity amplitudes and the squared
matrix element. We have checked that our code passed the
non--trivial test of electromagnetic gauge invariance.  We
employed {\tt Vegas} \cite{vegas} to perform the Monte Carlo
phase space integration with the appropriate cuts to obtain the
differential and total cross sections.

Our results were obtained assuming $\sqrt{s} = 500 \; \text{GeV}$
and an integrated luminosity  ${\cal L} = 50 \;\text{fb}^{-1}$.
We adopted a cut in the photon energy of $E_\gamma > 20 \;
\text{GeV}$ and required the angle between any two particles to
be larger than $15^\circ$. In these conditions,
$\sigma_{WW\gamma}^{SM} \simeq 144$ fb and we can expect around
$7200$ events per year. 

As we could expect, the $W$--pair production at NLC is able to
put a limit that is one order of magnitude better for the
coefficients $f_{B,W,WWW}$ \cite{hagiwara1}. However this
reaction is {\it not} able to constraint $f_{BB,WW}$.  This can
be accomplished using the reaction $e^+ e^- \to W^+ W^- \gamma$,
which gives rise to bounds on the values of these coefficients. 

\vskip 0.8cm
\begin{center}
\begin{tabular}{||c||c||c||}
\hline 
\hline
$M_H$(GeV) & $f_{BB}/\Lambda^2$ & $f_{WW}/\Lambda^2$ \\
\hline 
\hline
170 & ~~~~~ ( $-$11.8 , 6.6 )  ~~~~~ &  ~~~~~ ( $-$6.4 , 4.7 )  ~~~~~ \\
\hline
200 & ( $-$12.7 , 8.6 ) & ( $-$7.4 , 5.8 ) \\
\hline
250 & ( $-$14.1 , 11.3 ) & ( $-$8.5 , 7.7 ) \\ 
\hline
300 & ( $-$17.4 , 15.1 ) & ( $-$10.2 , 9.6 ) \\ 
\hline
350 & ( $-$24.0 , 21.5 ) & ( $-$15.5 , 13.7 ) \\
\hline 
\hline   
\end{tabular}

\vskip 0.5cm
{\it TABLE I. Range of the allowed values of the  coefficients
$f_{BB}$ and $f_{WW}$, in TeV$^{-2}$, for a $2\sigma$ deviation
in the total cross section of the process $e^+ e^- \to W^+ W^-
\gamma$ at NLC.}
\end{center}

The contribution of the anomalous couplings $f_{BB,WW}$ is
dominated by  on--mass--shell Higgs production with the
subsequent $H \to W^+ W^-$ decay.  Therefore the best constraints
are obtained at NLC for Higgs boson masses in the range $2 M_W
\leq M_H \leq (\sqrt{s} - E_\gamma^{\text{min}})$ GeV, {\it
i.e.\/} when  on--shell production is allowed. We present in
Table I the limits on the coefficients $f_{BB}$ and $f_{WW}$
based on a $2\sigma$ deviation in the total cross section for a
Higgs mass in the range  $170 \leq M_H \leq 350$ GeV. In Fig.\ 1
(a), we present the results of a combined sensitivity analysis in
the form of a contour plot for the two free parameter, $f_{BB}$
and $f_{WW}$, for $M_H = 170$ GeV. 

In an attempt to increase the sensitivity of this reaction, we
have considered the effect of electron beam polarization in
reducing the SM background. Even assuming a $90\%$ degree of
polarization for right--handed (RH) electrons, no significant 
improvement was obtained.

\begin{center}
\mbox{\epsfig{file=fig01.epsi,width=0.8\textwidth}}
\end{center}

\begin{center}
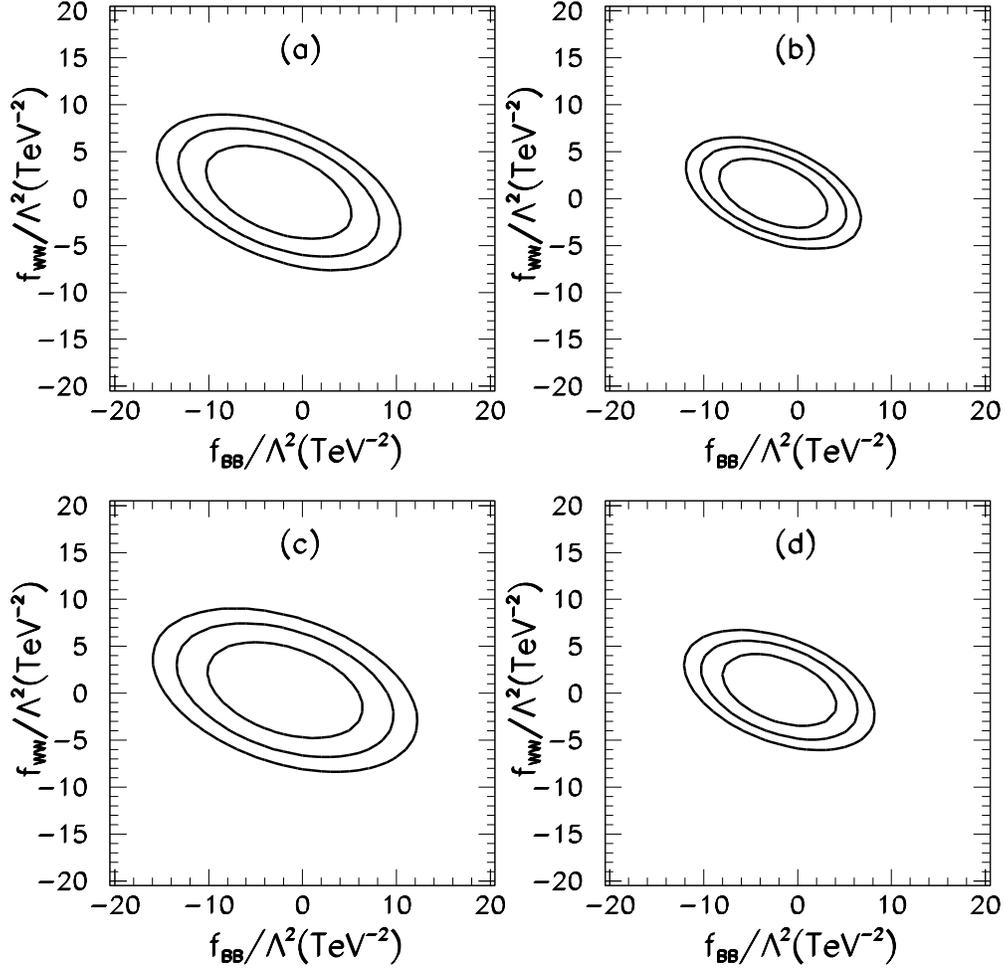
{\it FIGURE 1. Contour plot of $f_{BB} \times f_{WW}$.  The
curves show the 1, 2, and 3 $\sigma$ deviations from the SM total
cross section, for $e^+ e^- \to W^+ W^- \gamma$ ($M_H =170$ GeV),
with (a) no cut on $p_{T_\gamma}$, and  (b) cut of $p_{T_\gamma} >
100$ GeV, and for $e^+ e^- \to Z Z \gamma$ ($M_H =200$ GeV), with
(c) no cut on $p_{T_\gamma}$, and (d) cut of $p_{T_\gamma} > 100$
GeV.}
\end{center}

Since we expect the new interactions to involve mainly
longitudinally polarized gauge bosons, we studied the sensitivity
for different combinations of the polarizations of the $W$'s.
An analysis of the photon transverse momentum  distribution for
the different polarizations of the $W$'s shows the relevance   of
the $W_L W_L$ production for the anomalous contributions
\cite{us}. However, the improvement is very small because the
requirement of longitudinally polarized $W$ bosons reduces
drastically the total yield.

We have also investigated different distributions of the final
state particles in order to search for kinematical cuts that
could improve the NLC sensitivity. In the photon transverse
momentum we observe that  the contribution of the anomalous
couplings is larger in the high  $p_{T_\gamma}$ region.
Therefore, a cut of $p_{T_\gamma} > 100$ GeV drastically  reduces
the background. The improvement on $f_{WW,BB}$ bounds can be
clearly  seen from Fig.\ 1(b) where a 1, 2, and 3$\sigma$
deviations in the  total cross section, for $M_H=170$ GeV, is
shown after the above cut is  applied. 

\section{Anomalous $Z Z\gamma$ and $Z\gamma\gamma$
Production at NLC}

Through the processes $e^+ e^- \to ZZ\gamma$ and $e^+ e^- \to
Z\gamma\gamma$  we are also able to establish constraints on the
coefficients $f_{WW,BB}$ at NLC. Now there are 7 (10) standard
and 8 (12) anomalous Feynman diagrams involved in the
$Z\gamma\gamma$ ($Z Z \gamma$) production. These reactions 
were computed in the same way as before, and assuming the same
cuts {\it i.e.\/} $E_\gamma > 20$ GeV and the angle
between any two particles larger than $15^\circ$.

The reaction $e^+ e^- \to Z\gamma\gamma$ yields $2900$
events per year, assuming the expected luminosity for NLC
(${\cal L} = 50$ fb$^{-1}$). Requiring a maximum deviation of
2$\sigma$ in the total cross section, and  assuming a Higgs boson
of 200 GeV, we obtain the allowed ranges  $-39 < f_{BB}/\Lambda^2
< 35$ TeV$^{-2}$ and $-9.6 < f_{WW}/\Lambda^2 < 14$ TeV$^{-2}$.

The best constraint on the anomalous couplings come from the
reaction $e^+ e^- \to Z Z \gamma$, which gives rise to $\sim 675$
events per year at $\sqrt{s} = 500$ GeV. The contribution of the
anomalous couplings is also dominated by  on--mass--shell Higgs
production with the subsequent $H \to Z Z$ decay.  Therefore the
best results are obtained at NLC for Higgs boson masses  in the
range $2 M_Z \leq M_H \leq (\sqrt{s} - E_\gamma^{\text{min}})$
GeV, where on--shell production is allowed. We present in Table
II the limits on the coefficients $f_{BB}$ and $f_{WW}$ based on
a $2\sigma$ deviation in the total cross section for a Higgs mass
in the range  $200 \leq M_H \leq 350$ GeV.

\vskip 0.8cm
\begin{center}
\begin{tabular}{||c||c||c||}
\hline 
\hline
$M_H$(GeV) & $f_{BB}/\Lambda^2$ & $f_{WW}/\Lambda^2$ \\
\hline 
\hline
200 &  ~~~~~ ( $-$12.6 , 8.5 ) ~~~~~  &  ~~~~~ ( $-$6.6 , 5.7 ) ~~~~~  \\
\hline
250 & ( $-$13.3 , 9.5 ) & ( $-$7.5 , 6.2 ) \\ 
\hline
300 & ( $-$16.0 , 12.5 ) & ( $-$9.1 , 7.9 ) \\ 
\hline
350 & ( $-$21.9 , 18.3 ) & ( $-$12.3 , 11.5 )  \\
\hline 
\hline  
\end{tabular}

\vskip 0.5cm
{\it TABLE II. Range of the allowed values of the  coefficients
$f_{BB}$ and $f_{WW}$, in  TeV$^{-2}$, for a $2\sigma$ deviation
in the total cross section of the process $e^+ e^- \to Z Z
\gamma$ at NLC. }
\end{center}

We tried to take advantage of the fact that the new interactions
affect mostly the longitudinally polarized gauge bosons by
studying the polarized  $Z$ production. Unfortunately the
improvement obtained is very small since this requirement reduces
by two orders of magnitude  the total yield.

We have also investigated different distributions of the final
state particles in order to search for kinematical cuts that
could improve the NLC sensitivity. The most promising variable is
again the photon transverse momentum. Similarly to the $WW\gamma$
production, the contribution of the anomalous couplings is larger
in  the high $p_{T_\gamma}$ region, and a cut of $p_{T_\gamma} >
100$ GeV is able to reduce in a significant way the background.
The improvement on $f_{WW,BB}$ bounds can  be clearly seen from
Fig.\ 1 (c) and 1 (d) where the deviations in the total cross
section, for $M_H=200$ GeV, are shown before and after the above
cut is applied. 

The limits for $f_{BB}$ and $f_{WW}$ obtained from the study of
the reactions  $e^+ e^- \to W^+ W^- \gamma$ and $e^+ e^- \to Z Z
\gamma$, presented in Tables I and II, are comparable and both
processes  can be used as complementary tool in the study of
anomalous Higgs interactions.

\section{Conclusion}

The search for the effect of higher dimensional operators that
give rise to anomalous bosonic couplings may provide important
information on physics beyond the Standard Model. We have
analysed  the contributions of anomalous couplings arising from
dimension--6 operators of a linearly realized $SU_L(2) \times
U_Y(1)$ invariant effective Lagrangian in the processes $e^+ e^-
\to W^+ W^-  \gamma$, $e^+ e^- \to Z Z \gamma$, and $e^+ e^- \to
Z \gamma \gamma$.  We have included in our calculations, {\it
all} the anomalous trilinear and  quartic gauge couplings, the
anomalous Higgs couplings with gauge bosons and we have taken
into account the Standard Model irreducible background. The
impact of the anomalous contributions in the total cross section
of these processes were analysed at the NLC. Polarization of the
final state bosons are found to be insufficient to improve the
limits obtained from the total cross section of the three
processes. 

We mainly focused on the operators ${\cal O}_{WW}$ and  ${\cal
O}_{BB}$, which cannot be tested in the  $W-$pair production
process. We showed that a photon transverse momentum cut, 
$p_{T_\gamma} > 100$ GeV, can improve the limits on  $f_{WW}$ and
$f_{BB}$ obtained through an analysis of the total cross section.
Typical sensitivities of a few TeV$^{-2}$ are obtained at the NLC
for these coefficients.

\acknowledgments
This work was supported by Conselho Nacional de Desenvolvimento
Cient\'{\i}fico e Tecnol\'ogico (CNPq), and by Funda\c{c}\~ao de
Amparo \`a Pesquisa do Estado de S\~ao Paulo (FAPESP).



\begin{references}
\bibitem{foot} In our analysis, we have also taken into account
the SM one--loop Higgs contributions to the couplings
$H\gamma\gamma$ and $HZ\gamma$.
\end{references}
\end{document}